\documentclass[sigconf,authorversion]{acmart}
\usepackage{siunitx}
\usepackage{tabularx}
\newcommand{\tablefontsize}{\footnotesize}

\newcommand{\tableheadline}{\textbf}

\usepackage{pifont}

\newcommand{\yes}{\ding{51}}

\usepackage[english]{babel} 
\usepackage{babel}
\addto\extrasenglish{%

}
\microtypecontext{spacing=nonfrench}

\usepackage[acronym,nohypertypes={acronym},shortcuts,nonumberlist,nolist]{glossaries} 
\newacronym{E2EE}{E2EE}{end-to-end encrypted}
\newacronym[description={Machine-in-the-Middle}]{MitM}{MitM}{machine-in-the-middle}
\newacronym[description={Near-Field Communication}]{NFC}{NFC}{near-field communication}
\newacronym[description={Out-Of-Band}]{OOB}{OOB}{out-of-band}
\newacronym[description={Quick-Response}]{QR}{QR}{quick-response}

\AtBeginDocument{%
  \providecommand\BibTeX{{%
    \normalfont B\kern-0.5em{\scshape i\kern-0.25em b}\kern-0.8em\TeX}}}

\microtypecontext{spacing=nonfrench}

\copyrightyear{2024}
\acmYear{2024}
\setcopyright{acmlicensed}
\acmConference[CSCW Companion '24] {Companion of the 2024 Computer-Supported Cooperative Work and Social Computing}{November 9--13, 2024}{San Jose, Costa Rica.}
\acmBooktitle{Companion of the 2024 Computer-Supported Cooperative Work and Social Computing (CSCW Companion '24), November 9--13, 2024, San Jose, Costa Rica}
\acmISBN{979-8-4007-1114-5/24/11} 
\acmDOI{10.1145/3678884.3681818}

\acmSubmissionID{1028}

\begin{document}

\title[PairSonic: Helping Groups Securely Exchange Contact Information]{PairSonic: \\Helping Groups Securely Exchange Contact Information}

\author{Florentin Putz}
\orcid{0000-0003-3122-7315}
\affiliation{%
  \institution{Technical University of Darmstadt}
  \city{Darmstadt}
  \country{Germany}
  }
\email{fputz@seemoo.de}
  
\author{Steffen Haesler}
\orcid{0000-0002-6808-0487}
\affiliation{%
  \institution{Technical University of Darmstadt}
  \city{Darmstadt}
  \country{Germany}
}
\email{haesler@peasec.tu-darmstadt.de}

\author{Thomas Völkl}
\orcid{0009-0004-1051-1549}
\affiliation{%
  \institution{Technical University of Darmstadt}
  \city{Darmstadt}
  \country{Germany}
}
\email{tvoelkl@seemoo.de}

\author{Maximilian Gehring}
\orcid{0009-0004-5599-5807}
\affiliation{%
  \institution{Technical University of Darmstadt}
  \city{Darmstadt}
  \country{Germany}
}
\email{gehring@cs.tu-darmstadt.de}

\author{Nils Rollshausen}
\orcid{0000-0003-2445-8684}
\affiliation{%
  \institution{Technical University of Darmstadt}
  \city{Darmstadt}
  \country{Germany}
}
\email{nrollshausen@seemoo.de}

\author{Matthias Hollick}
\orcid{0000-0002-9163-5989}
\affiliation{%
  \institution{Technical University of Darmstadt}
  \city{Darmstadt}
  \country{Germany}
}
\email{mhollick@seemoo.de}

\renewcommand{\shortauthors}{Florentin Putz et al.}

\begin{abstract}
Securely exchanging contact information is essential for establishing trustworthy communication channels that facilitate effective online collaboration.
However, current methods are neither user-friendly nor scalable for large groups of users.
In response, we introduce PairSonic, a novel group pairing protocol that extends trust from physical encounters to online communication.
PairSonic simplifies the pairing process by automating the tedious verification tasks of previous methods through an acoustic out-of-band channel using smartphones' built-in hardware.
Our protocol not only facilitates connecting users for computer-supported collaboration, but also provides a more user-friendly and scalable solution to the authentication ceremonies currently used in end-to-end encrypted messengers like Signal or WhatsApp.
PairSonic is available as open-source software: \url{https://github.com/seemoo-lab/pairsonic}
\end{abstract}

\ccsdesc[500]{Human-centered computing~Computer supported cooperative work}
\ccsdesc[300]{Human-centered computing~Open source software}
\ccsdesc[500]{Human-centered computing~Collaborative interaction}
\ccsdesc[500]{Human-centered computing~Smartphones}
\ccsdesc[100]{Human-centered computing~Mobile computing}
\ccsdesc[500]{Security and privacy~Authentication}
\ccsdesc[300]{Security and privacy~Usability in security and privacy}

\keywords{usable security; group pairing; secure device pairing; trust establishment; acoustic communication; data-over-sound;
authentication ceremony;
public-key cryptography; end-to-end encryption; instant messaging; online collaboration; contact management;
key verification; peer-to-peer-authentication; MitM attacks;
device association; binding; bonding; coupling;
smartphone; prototype; security; privacy; social cybersecurity; interoperability; ad-hoc; decentral; spontaneous; nearby; proximity; open-source; WiFi; NFC; Android
}

\begin{teaserfigure}
\centering
\includegraphics[width=\textwidth]{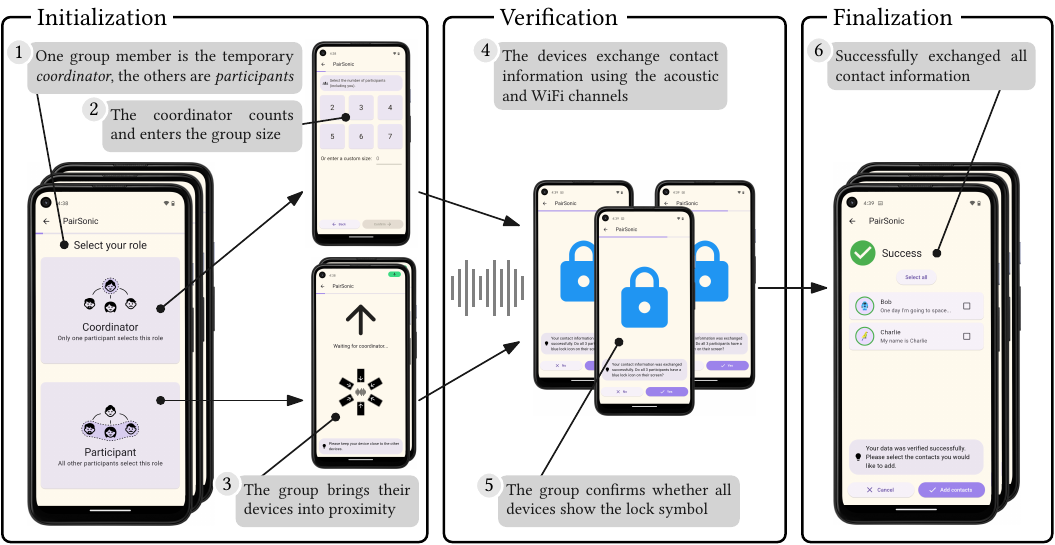}
	\caption{%
    The PairSonic user interface, demonstrated for a group of three users wanting to exchange their contact information.
	}
        \Description{%
		    An image showing three boxes labeled "Initialization", "Verification", and "Finalization", for each of the three protocol phases of PairSonic. Each box contains screenshots of the PairSonic app inside a smartphone frame.
            The first screenshot in in the "Initialization" box and shows the role selection and is labeled with the following text:  "One group member is the temporary coordinator, the others are participants.". An arrow points from the "Coordinator" button in the screenshot to a second screen within the "Initialization" box, which shows a slider to select a number. This second screen is labeled with the following text: "The coordinator counts as enters the group size".
            Another arrow points from the "Participant" button in the first screenshot to a third screen within the "Initialization" box, which prompts the users to align their devices close to each other. This screen is labeled with the following text: "The group brings their devices into proximity".
            The second and third screen are annotated with an arrow pointing to a fourth screen, which is inside the "Verification" box. The box is labeled with the following text: "The devices exchange contact information using the acoustic and WiFi channels".
            The fourth screen shows a blue lock symbol and is labeled with the following text: "The group confirms whether all devices show the lock symbol".
            An arrow from this screen points towards a fifth screen in the "Finalization" box, which shows the text "Success" and the newly exchanged contacts. This screen is labeled with the following text: "Successfully exchanged all contact information".
		}
	\label{fig:overview}
\end{teaserfigure}

\maketitle
\section{Introduction}

Effective communication and collaboration require a trusted environment where participants clearly know each other's identities and do not have to fear unauthorized access~\cite{oesch2022User}.
This trust is vital for facilitating open discussions among friends, family, colleagues, business partners, and especially vulnerable groups like journalists and activists~\cite{shusas2023accounting, fassl2023Why,namara2021differential}.
While \gls{E2EE} messengers like Signal support group collaboration through real-time messaging and video calls, they cannot, by default, protect against unauthorized access and impersonation by active adversaries.
These threats include rogue service operators and \gls{MitM} attacks.
Active \gls{MitM} attacks in particular have become more prominent due to reports of law enforcement agencies compromising app providers, highlighting a practical and concerning risk for privacy infringements by malicious actors in the future~\cite{fassl2023Why}.

To mitigate these risks, \gls{E2EE} tools require users to perform a manual contact verification process known as the \textit{authentication ceremony}~\cite{vaziripour2018Action}.
Only after this ceremony is completed is the communication fully protected.
The current design of authentication ceremonies in \gls{E2EE} tools has two major shortcomings:
First, they have low usability, requiring considerable user interaction, time, and effort~\cite{oesch2022User,fassl2023Why,vaziripour2018Action}.
Second, they do not support larger groups, functioning only between two users~\cite{cscw2024pairsonic}.
This limitation is significant because collaboration often involves multiple participants.
Performing bilateral authentication ceremonies between each pair of group members is infeasible, as the number of required manual verifications scales quadratically with the group size.

To address both challenges, we designed PairSonic, a novel pairing protocol for the fast and secure exchange of contact information that scales efficiently to multiple participants.
Our prototype implementation allows users to extend trust from physical encounters to online communication by simply holding their devices together for a few seconds, aligning with intuitive association behavior.
This demo accompanies our full paper at CSCW~2024~\cite{cscw2024pairsonic}, where we evaluated PairSonic's usability in a lab study, finding it to be highly effective for groups.
PairSonic automates the tedious verification tasks required by previous approaches, greatly enhancing usability and making secure group communication more accessible.

\section{System Design}
PairSonic is a group pairing protocol that enables two or more users meeting in person to spontaneously exchange or verify their contact information.
Each participant obtains authentic and verified contact details of all other participants, including their cryptographic public keys.
Importantly, our protocol does not rely on external key management infrastructure, prior associations, or shared secrets.

\subsection{Protocol}
The PairSonic protocol is based on the cryptographic foundation of the SafeSlinger protocol~\cite{farb2013SafeSlinger}, but improves its usability and deployability~\cite{cscw2024pairsonic}. Our protocol operates in three main phases (\autoref{fig:overview}):

\begin{enumerate}
    \item \textbf{Initialization:} A group of users initiates PairSonic on their smartphones to exchange contact data. One member is selected as the temporary \textit{coordinator} while the others become \textit{participants}. The coordinator inputs the total group size and ensures that all participants' smartphones are ready and within proximity.
    The coordinator's smartphone sets up a temporary ad-hoc WiFi network and transmits the network details over an acoustic \gls{OOB} channel, allowing the other smartphones to connect automatically.

    \item \textbf{Verification:}
    Once connected, the devices use this WiFi network to perform a cryptographic protocol derived from the SafeSlinger protocol~\cite{farb2013SafeSlinger}.
    Participants send nested commitments, containing their contact data and cryptographic public keys, to the coordinator's smartphone. The coordinator aggregates these commitments into a hash value and transmits it over the acoustic \gls{OOB} channel. All devices monitor this channel, ready to abort the process if an incorrect message is detected.

    Upon verifying the correct hash value, a blue lock symbol appears on each smartphone, prompting users to confirm the matching symbols on all other devices.
    If confirmed, the smartphones use the WiFi network to distribute success nonces, allowing the display of verified exchanged contacts. If the lock symbol confirmation fails, abort nonces are released, halting the protocol and ensuring the integrity of the contact exchange.
    We changed the verification symbol based on feedback from our user study, because some participants misunderstood the previous checkmark as indicating protocol completion without verifying the symbol.

    \item \textbf{Finalization:}
    Each participant now sees a list of all received contacts and can select which to import.
\end{enumerate}

\begin{table*}
\caption{Selection of devices where we have verified compatibility with PairSonic, indicating support for audible and inaudible modes. In principle, any smartphone with WiFi Direct and running at least Android 6.0 (released in 2015) should be compatible.}
\label{tab:compatibility}
\begin{minipage}[t]{.475\linewidth}
    \vspace{0pt}
    \centering
    \tablefontsize
    \begin{tabularx}{\columnwidth}{lllll}
    \toprule
    \tableheadline{Model} &
    \tableheadline{OS} &
    \tableheadline{SDK} &
    \tableheadline{Audible} &
    \tableheadline{Inaudible}\\
    \midrule
    Google Pixel 4 & Android 13 & 33 & \yes & \yes \\
    Google Pixel 4a & Android 13 & 33 & \yes & \yes \\
    Google Pixel 5 & Android 13 & 33 & \yes & \yes \\
    Google Pixel 6 Pro & Android 13 & 33 & \yes & \yes \\
    LG Nexus 5 & Android 6.0.1 & 23 & \yes & \yes \\
    Motorola Edge & Android 10 & 29 & \yes & \yes \\
    OnePlus 10 Pro & Android 12 & 31 & \yes & \yes \\
    \bottomrule
    \end{tabularx}
\end{minipage}\hfill
\begin{minipage}[t]{.475\linewidth}
    \vspace{0pt}
    \centering
    \tablefontsize
    \begin{tabularx}{\columnwidth}{lllll}
    \toprule
    \tableheadline{Model} &
    \tableheadline{OS} &
    \tableheadline{SDK} &
    \tableheadline{Audible} &
    \tableheadline{Inaudible}\\
    \midrule
    Oppo Reno 6 & Android 11 & 30 & \yes & \yes \\
    Redmi Note 9 Pro & Android 10 & 29 & \yes & \yes \\
    Samsung Galaxy A32 & Android 12 & 31 & \yes & \yes \\
    Samsung Galaxy A52s 5G & Android 14 & 34 & \yes & \yes \\
    Samsung Galaxy S10e & Android 12 & 31 & \yes & \yes \\
    Samsung Galaxy S20 Ultra 5G & Android 10 & 29 & \yes & \yes \\
    Samsung Galaxy S22 & Android 13 & 33 & \yes & \yes \\
    \bottomrule
    \end{tabularx}
\end{minipage} 
\end{table*}

\subsection{Acoustic Out-Of-Band Channel}
Previous user studies have indicated that performing authentication ceremonies with current \gls{E2EE} tools like Signal or WhatsApp is often challenging due to the time and effort required~\cite{oesch2022User,fassl2023Why,vaziripour2018Action}.
Our protocol, PairSonic, addresses these issues by automating manual verification tasks such as comparing phrases, digits, or pictures.
PairSonic utilizes an acoustic location-limited channel, leveraging the smartphones' built-in speakers and microphones to exchange data within a range of approximately one meter.
This method simplifies the inclusion of multiple devices at once, a distinct advantage over \gls{NFC} or \gls{QR} codes, which typically handle one device at a time.
Moreover, unlike WiFi or Bluetooth, the acoustic physical layer in our system is inherently location-limited and software-defined, allowing for the incorporation of advanced physical layer security techniques~\cite{putz2020AcousticIntegrityCodes}.
These enhancements can significantly strengthen the protocol's defense against sophisticated attacks on the acoustic channel, thus providing an additional layer of security for group pairing.

\section{PairSonic Prototype}
Our prototype of PairSonic is implemented as an Android app using the Flutter framework. For communication, we utilize two channels: a local ad-hoc WiFi connection, provided by the Android WiFi Direct API, and the acoustic \gls{OOB} channel for data transmission via sound waves, provided by the ggwave library.\footnote{Data-over-sound library ggwave: \url{https://github.com/ggerganov/ggwave}}
During our user study~\cite{cscw2024pairsonic}, we employed audible frequencies ranging from \SI{1875}{\hertz} to \SI{6375}{\hertz} to evaluate user perceptions of this communication method.
Participants expressed concerns about the appropriateness of audible noises in quiet settings like museums or libraries.
In response, we developed a variant that uses inaudible near-ultrasonic frequencies between \SI{15000}{\hertz} and \SI{19500}{\hertz}.
Our testing revealed that these inaudible frequencies not only mitigate the issue of noise in quiet environments but also enhance the robustness of the communication.
This improvement addresses another concern raised in our lab study, which noted occasional failures in acoustic communication that required restarting the pairing process.

The combination of an acoustic channel with WiFi Direct in our prototype forms a novel communication stack. To ensure functionality and compatibility, we conducted tests using smartphones from various manufacturers (\autoref{tab:compatibility}). The prototype functioned effectively on all devices tested that support WiFi Direct, provided they run at least Android version 6.0 (API level 23, released in 2015).
PairSonic is interoperable, meaning that different device models within the exchange can interact seamlessly.

\section{Applications}
PairSonic is designed to assist users in establishing trustworthy digital communication channels, making it especially useful for forming ad-hoc collaboration groups among multiple users~\cite{namara2021differential}. 
This makes PairSonic particularly adaptable to various computer-supported collaboration tools, including group work platforms, educational hubs, and video conferencing systems.
Furthermore, our protocol could be integrated into E2EE messengers like Signal, WhatsApp, or Matrix, providing a more user-friendly authentication ceremony to support effective and trustworthy online collaboration.

Our study~\cite{cscw2024pairsonic} highlights these potential applications: participants identified multiple scenarios where PairSonic could serve as a secure method for contact exchange. These include forming private chat groups on Signal, managing participants in Zoom video conferences, and enhancing interaction within online collaboration tools.

\section*{Availability}
Together with this paper, we provide an overview of the PairSonic project online at \url{https://seemoo.de/s/pairsonic}.
PairSonic is available as open-source software on GitHub:  \url{https://github.com/seemoo-lab/pairsonic}

\begin{acks}
This work has been funded by the
LOEWE initiative (Hesse, Germany)
within the emergenCITY center\newline
[LOEWE/1/12/519/03/05.001(0016)/72].
\end{acks}

\bibliographystyle{ACM-Reference-Format}
\typeout{} 
\bibliography{bibliography}

\end{document}